\documentclass[aps,pre,groupedaddress]{revtex4-1}

\usepackage[latin9]{inputenc}
\usepackage{amsmath}

\makeatletter

\usepackage{graphicx}
\graphicspath{ {images/} }
\usepackage[caption=false]{subfig}
\usepackage{float}
\usepackage{hyperref}

\makeatother

\usepackage{adjustbox}

\usepackage[normalem]{ulem}

\begin{document}
\linespread{1.61}\selectfont{}

\title{Dynamical clustering of U.S. states reveals four distinct infection patterns that predict SARS-CoV-2 pandemic behavior}

\author{Joseph L. Natale}
\affiliation{Hal{\i}c{\i}o\u{g}lu Data Science Institute, University of California, San Diego, La Jolla, CA 90293, USA}

\author{Varun Viswanath}
\affiliation{Department of Electrical and Computer Engineering, University of California, San Diego, La Jolla, CA 90293, USA}

\author{Oscar Trujillo Acevedo}
\affiliation{Divisi\'{o}n de Ingenier\'{i}a Bioqu\'{i}mica, Instituto Tecnol\'{o}gico Superior of Atlixco, C. Heliotropo No.1201, Unidad 8 Norte Nueva Xalpatlaco, Vista Hermosa, 74210 Atlixco, Pue.}

\author{Sophia P\'{e}rez Giottonini}
\affiliation{Colegio de Bachilleres del Estado de Sonora, plantel Villa de Seris, Hermosillo, SON, M\'{e}xico.}

\author{Sandy Ihuiyan Romero Hern\'{a}ndez}
\affiliation{Instituto Tecnol\'{o}gico Superior de Atlixco, C. Heliotropo No.1201, Unidad 8 Norte Nueva Xalpatlaco, Vista Hermosa, 74210 Atlixco, Pue., M\'{e}xico}

\author{Diana G. Cruz Mill\'{a}n}
\affiliation{Department of Health Sciences, Universidad de Sonora, campus Cajeme, Cd. Obreg\'{o}n, Son., M\'{e}xico}

\author{A. Montserrat Palacios-Puga}
\affiliation{Department of Biochemical and Chemical Engineering, Instituto Tecnol\'{o}gico de Tijuana, Tecnol\'{o}gico Nacional de M\'{e}xico, Tijuana, B.C, M\'{e}xico}

\author{Ammar Mandvi}
\affiliation{Department of Emergency Medicine, Department of Family Medicine, University of California, San Diego, CA, USA}

\author{Brian M. Khan}
\affiliation{Department of Emergency Medicine, Department of Internal Medicine, University of California, San Diego, CA, USA} 

\author{Martin Lilik}
\affiliation{Department of Software Engineering, Park Media, Semarang, Indonesia}

\author{Jay Park}
\affiliation{Department of Emergency Medicine, University of California, San Diego, La Jolla, CA 92037}

\author{Benjamin L. Smarr}
\affiliation{Department of Bioengineering and Hal{\i}c{\i}o\u{g}lu Data Science Institute, University of California, San Diego, La Jolla, CA 90293, USA}

\date{November 19, 2021}

\begin{abstract}

The SARS-CoV-2 pandemic has so far unfolded diversely across the fifty United States of America, reflected both in different time progressions of infection ``waves" and in magnitudes of local infection rates. Despite a marked diversity of presentations, most U.S. states experienced their single greatest surge in daily new cases during the transition from Fall 2020 to Winter 2021. Popular media also cite additional similarities between states -- often despite disparities in governmental policies, reported mask-wearing compliance rates, and vaccination percentages. Here, we identify a set of robust, low-dimensional clusters that 1) summarize the timings and relative heights of four historical COVID-19 ``wave opportunities" accessible to all 50 U.S. states, 2) correlate with geographical and intervention patterns associated with those groups of states they encompass, and 3) predict aspects of the ``fifth wave" of new infections in the late Summer of 2021. In particular, we argue that clustering elucidates a negative relationship between vaccination rates and subsequent case-load variabilities within state groups. We advance the hypothesis that vaccination acts as a ``seat belt," in effect constraining the likely \emph{range} of new-case upticks, even in the context of the Summer 2021, variant-driven surge.

\end{abstract}


\maketitle 

\section{Introduction}

The SARS-CoV-2 pandemic has claimed millions of lives globally -- and directly affected hundreds of millions more -- doing so in a highly unequal manner~\cite{van2020covid,chaudhry2020country}. Early foci for COVID-19, such as northern Italy, coincided with populations with high proportions of elderly, susceptible people~\cite{ciminelli2020covid}; the availability of hospital beds, treatments, and clinical personnel are again to be a major problem in many geographical areas~\cite{weissman2020locally,sen2021closer}; and new variants are emerging at different rates in different locales~\cite{alfaro2021closer,chadha2021facing}.

Internal to the United States of America (at the time of writing, that nation with the highest recorded numbers of both ``total" COVID-19 cases, and COVID-19-related deaths) disparities also permeate local infection rates~\cite{loomba2021disparities,hamman2021disparities}. This has prompted a number of explicit comparisons, at the state level~\cite{xian2020racial,solis2021understanding}. In principle, such collations should be of great interest to public health decision-makers: to the extent that different states can be expected to behave in categorically different ways, information that supports early planning for interventional responses can be gleaned by mapping the states' diverse progressions over time.

Publicly available data render unambiguous that all fifty U.S. states saw their greatest, sustained incidence of new cases during the country's last winter season (2020 - 2021)~\cite{centers2021covid}. Furthermore, media outlets have directed attention to other, ostensible similarities: in the case of Florida and California, news coverage emphasized an attainment of ``similar COVID-19 case rates"~\cite{healthline}, or even allegedly ``identical outcomes"~\cite{identical_outcomes}, despite the drastically different measures taken by the independently-operating, state-level governments at the time of reporting. Still, in complex systems it is always possible that multiple parts or subsystems exhibit temporarily superficial similarities without sharing their underlying dynamics~\cite{strogatz_book}. Whether or not the U.S. state infection histories fall into universal similarity classes with relevance to pandemic planning -- and, indeed, questions of how reliably government policy, and even vaccination rates, determine their ultimate case statistics -- remains unanswered by these prior analyses.

In this manuscript we consider one specific approach to demonstrate that the historical ``infection trajectories" for all fifty U.S. states can be grouped according to some reasonable measure of similarity, such that resulting groups form a low-dimensional set (i.e., one much smaller than the number of individual states), and also prove useful for predicting the present, or even future, of this pandemic -- not merely its past. As such groupings are established, any relevant information supporting early planning around responses to the states' diverse presentations can be contextualized and made available to public health decision-makers.

\section{Methods}
\label{sec:2}

\subsection{Data Acquisition and Preparation}
\label{sec:2a}

\subsubsection{Infection Data}
\label{inf_data}

The data referenced in the subsequent paragraphs were aggregated from sources developed by Johns Hopkins and Oxford Universities. The infection data, reported at daily resolution, were obtained from Johns Hopkins~\cite{jhucrc}. These included reports of \emph{total} (confirmed and probable) \emph{cases}, expressed as cumulative counts for each U.S. state. In order to transform these cumulative records into ``daily new" infection counts for each state, we subtracted each day's \emph{total cases} record from the previous day's value, and then replaced their raw counts with the corresponding moving (14-rolling) averages attained by each day considered. In so doing, we therefore neglected the effects of deaths and recoveries on the ``daily new" case numbers; we justified this approximation by noting that a) the latter were often as small as a full order of magnitude less than the total cases on a given day, and b) isolated ``negative" days (those for which deaths and recoveries presumably exceeded the new, emerging case count) would be absorbed, without a strong effect, by the aforementioned temporal averaging. Training data were amassed between April 2020 to July 2021.

Since the different U.S. states encompass populations with markedly different sizes and densities, but our temporal picture seeks to distill all the states' histories into a form for which their synchrony and structure can be compared on equal footing, we normalized the aforementioned, raw estimates of the states' daily new infection counts by their respective maximum values (all observed to occur in the same time frame, Winter 2020 - 2021). This yielded a new set of case records for every state, with possible values between 0 and 1. Treating each as a time series, and applying 14-day moving (rolling) averages between April 13, 2020 and July 2, 2021 -- a total of 446 calendar days -- defined our U.S. ``daily new" \emph{infection trajectories}. An example trajectory, for New York state, is shown in Fig.~\ref{fig:trajectorysamples}, in blue.

Daily new infection records for dates prior to April 13, 2020 are depicted for temporal context in Section~\ref{sec:3}, but were not used for developing or training any models: in these early stages of the pandemic, we could not be confident that the reporting standards across states were of uniform or sufficiently high quality~\cite{wu2020substantial}. For instance, a lack of PCR testing supplies might have led to a greater proportion of ``probable," vs. ``confirmed," cases so early on. Some numerical artifacts resulting from performing 14-day averages on data that begins midway through the initial surge of cases in Spring 2020 were removed from the figures in Section~\ref{sec:3}, by adjusting the plotting limits across the respective x-axes. During the course of this work, data were re-scraped to extend between March 6, 2020 and August 3, 2021.

\subsubsection{Definition of five ``wave opportunities"}
\label{def_waveopps}

The example infection trajectory for New York (in Fig.~\ref{fig:trajectorysamples}) includes periods of ``low" relative case loads, punctuated by several dramatic -- and, persistent -- elevations. A central question in the present work is whether the temporal patterns associated with these periodic surges in the daily new cases fall into stereotypes across states. As a starting point for comparing the time progressions realized by different states, we modify a previously established convention, dividing the pandemic's history between the Spring of 2020 and Spring 2021 into consecutive ``waves"~\cite{policyRef}. Here, we reconceptualize these divisions in terms of annual, calendar seasons:

\begin{enumerate}
    \item \textbf{First wave}: an initial case accumulation, roughly between March and May (``Spring") of 2020
    \item \textbf{Second wave}: a subsequent resurgence, roughly between July and September (``Summer") of 2020
    \item \textbf{Third wave}: the aforementioned ``Winter peak," lasting from around November 2020 through March 2021, but reaching its apex between December 2020 - January 2021
    \item \textbf{Fourth wave}: recalcitrant surge during the Spring of 2021 that was noted in, but had not necessarily resolved by the time of publication of,
    Ref.~\cite{policyRef}
\end{enumerate}

In addition, we consider a ``fifth" wave, encompassing more recent case load rises beginning in late July 2021 (Fig.~\ref{fig:trajectorysamples}). This upswing was reflected in the daily new case records for all fifty U.S. states, but the same was not universally true of the previous four waves. To emphasize explicitly that not all states participated equally in each of those four, we adopt the nomenclature \emph{wave opportunity} to refer to the potentially-eventful spans of time with persistent elevations. As one example, New York expressed strongly the ``first" wave opportunity, but the ``second wave" seen in other parts of the country, as defined above, was so small relative to the former that New York effectively progressed without participation in it. To describe the climbing infection rates in the Winter months, we refer here not to ``New York's second wave," but to ``New York's (degree of) participation in the third U.S. wave opportunity" (see again Fig.~\ref{fig:trajectorysamples}).

Since we are excluding data before April 13, 2020, and our rolling averages are fully-defined only after fourteen points have been observed (i.e., from April 26, 2020 on), the first wave opportunity is not considered in our analyses.

\subsubsection{Ancillary Data Sources}
\label{ancillary}

In addition to infection data, we aggregated additional data on vaccine administration rates and government policy ``stringencies" during the pandemic. We obtained vaccination data from repositories published by the Johns Hopkins University Coronavirus Resource Center \cite{jhucrc}, and curated by Our World in Data~\cite{owid_covid}. We incorporated government policy data compiled by the Blavatnik School of Government, University of Oxford~\cite{policyRef}.

The percentage of ``fully-vaccinated" individuals residing in each state -- our chosen proxy for progress toward a local ``herd immunity" status -- was estimated by, first, extracting the number of people recorded as having received \emph{two doses of either two-series} (Pfizer-BioNTech or Moderna), or \emph{one dose of the single-shot} (J\&J/Jannsen), vaccine against COVID-19 asvailable in the U.S., \emph{at each calendar day} between February 1, 2021 and July 22, 2021, inclusive. This time range, distinct from (but still overlapping with) that of our infection data, consisted of 172 days.

We aggregated tabulated counts by state, and then divided by the number of residents in each state. Although U.S. state population levels do change over time, we made the tacit approximation of slow change over the course of the pandemic: dividing by the ``2021" population counts given in Ref.~\cite{population2021} allowed us to arrive at a cumulative percentage of fully-vaccinated residents in each state as a time series -- a \emph{vaccination-rate} analog of our infection trajectories. The vaccination-rate trajectory for New York is presented as a red curve, normalized between 0 and 1, in Fig.~\ref{fig:trajectorysamples}.

A ``Stringency Index" score for comparing government policies across states was developed by previous authors, to summarize the ``strictness of `lockdown style' closure and containment policies that primarily restrict people's behavior"~\cite{policyRef}. Stringency scores could range, in principle, from a minimum value of 0 to a maximum of 100 for a given state; rescaling the stringency time series values provided a third, government-\emph{policy trajectory}, normalized to the same vertical scale as the infection and vaccine trajectories (between 0 and 1). Policy trajectories began much earlier than daily-infection and vaccination-rate trajectories, on January 1, 2020 (before the ``start" of the pandemic in the U.S.) and ended on July 2, 2021 -- a total of 549 days. A gray curve denotes the policy trajectory in Fig.~\ref{fig:trajectorysamples}.

\subsection{Correlation-Based Clustering of Infection Data}
\label{sec:2b}

Given the above considerations, we hypothesized that \textbf{i}) if placed on some even footing, the $n=50$ U.S. ``\emph{daily new infection}" \emph{trajectories} should collapse into a set $\mathcal{C}$ of archetypal forms with cardinality $|\mathcal{C}| \ll n$. That is, we set out to verify whether the records for all 50 states could be summarized by -- and compressed into -- a handful of discrete \emph{clusters}, each associated with a particular timing and structure (relative peak heights) for all those distinct ``wave opportunities" encompassed by the available data, between April 2020 and July 2021. We then investigated \textbf{ii}) whether one such partitioning might prove useful, not only for summarizing the states' dynamics in the \emph{past}, but also for predicting aspects of the \emph{subsequent}, ongoing surge, attributed largely to the B.1.617.2 (``Delta") variant~\cite{allen2021household}.

``Daily new infection" trajectory values spanning the time range between April 12th, 2020 and July 2nd, 2021 were clustered in Python 3, using the \emph{SciPy} Clustering package. Specifically, we performed hierarchical / agglomerative clustering via the named ``\emph{linkage}" function in the \emph{hierarchy} module of \emph{SciPy} (v. 1.4.1). A total of 446 data points per state, each representing a trajectory value for 1 calendar date, served as primary input to this function. We used a ``complete" (i.e., Farthest Point, or Voor Hees, Algorithm) linkage method to compute cluster distances. In what follows, this distance was measured by the (Pearson) \emph{correlation} (see~\cite{pdist}) between respective pairs of time series.

Using the named ``\emph{fcluster}" function of the same \emph{SciPy} module, we obtained the set of ``flat" clusters that would result from ``slicing" the discovered hierarchy at a specific \emph{threshold} value of our correlation-based distance metric. This same process could be repeated to cluster other types of data -- e.g., percent-fully-vaccinated ``trajectories".

\subsection{Principal Components Analysis (PCA)}
\label{sec:2c}

\subsubsection{Discovering ``important" date ranges that set states apart}

Principal Components Analysis was applied directly to the infection trajectories, treating them as 50 observations of one dynamical system, and the 446 daily new case values as their basis ``features" (those to be linearly combined). This was accomplished using the \emph{decomposition} module of the \emph{scikit-learn} package (v. 0.22.2.post1) in Python 3. Using the named function ``\emph{pca}," we extracted the first 50 principal components. Since these principal components are essentially linear combinations of the input features, each represents a weighted sum of contributions from the rolling, ``daily new infections" on different \emph{dates}; therefore, we used their feature importance rankings to reveal the subsets of \emph{dates} during the pandemic associated with the greatest degrees of variation across all 50 U.S. states. 

Feature importance, in the context of one given PCA-derived eigenvector (or, ``principal component"), was scored according to absolute values: the vector components with highest (relative) absolute values were ranked as the most important. In order to arrive at a more all-encompassing set of dates that took into account the importances from the first \emph{several} principal components $k \in \lbrace 1, \ldots k_0 \rbrace$, we also took a step back from the usual abstraction of principal components as (ideally, orthogonal) eigenvectors, and treated each instead as a set of general `importance-amplitude" vectors, having components for each of the 446 dates. Summing these importance-amplitude vectors, component-wise, resulted in a new curve that we used to represent ``overall," daily importances.

In order to select an appropriate value for $k_0$, we examined the cumulative percentage of ``explained variance" associated with the first several principal components $k \in \lbrace 1, 2, \dots 10 \rbrace$; these values are reported in Table I, along with the ``errors" corresponding to reconstructions of the 50 original observations -- our trajectories -- that were built using all principal components from 1 to $k_0$, inclusive. The ``error" at each time point was measured via the mean absolute deviation between the original and reconstructed trajectory value, averaged over states.  

The cutoff for choosing $k_0$ in terms of the cumulatively explained variance was set at a value of $90 \%$. In the context of the ``overall" scheme above, dates were considered ``important" if the absolute value of a given cumulative importance curve exceeded the mean value of that curve. An upper limit of $k=50$ (the minimum between the number of observations, 50, and the number of features, 466) principal components was permitted by the ``\emph{pca}" function.

The results for the most important date ranges inferred via PCA are depicted in Section IVc, contextualized by their superposition over all $n=50$ original state trajectories, with a common axis. There is no intrinsic guarantee that PCA should discover date ranges that are contiguous, for a generic time series decomposed in this manner.

\subsubsection{Developing ``minimal models" for U.S. state trajectories}
\label{master_dev}

In the course of performing trajectory reconstructions, we verified that those reconstructions built using just the first, $k=1$ principal component collapsed to a single curve. In other words, PCA lead to a description of the system that was identical  and equivalent for all $n$ states, at this ``coarsest" level of resolution. Isolating just the most prominent peaks within this curve (``find\textunderscore peaks" function, signal processing toolbox, \emph{SciPy} v.1.4.1, with the ``prominence" parameter set to the minimum-observed value of that single, $k=1$ reconstruction curve, 0.05) allowed us to \emph{compress} our ``$k=1$ reconstruction" into a handful of values representing its most pronounced excursions above its minimum. The dates associated with these maxima were then cross-referenced with those of the greatest overall ``importance," according to the first $k_0$ components cumulatively (see Sec.~\ref{mastermodel}).

Since the $k=1$, reconstructed trajectory represented a single, unifying summary of all $n$ original trajectories, we referred to this curve as the ``master model." The master model served as our \emph{minimal} model of the archetypal daily new infection trajectory, for an arbitrary U.S. state.

We quantified the errors associated with assuming that all states followed a given master-model behavior exactly. These included \emph{i)} the extent to which the full, $n=50$ master model failed to capture nuances in the $n$ individual trajectories from which it was inferred, and \emph{ii)} how well the aforedescribed compression, consisting exclusively of the master model \emph{maxima}, captured information conveyed by the master itself. To do so, we computed the mean absolute deviations between \emph{i)} the 446-day, $n=50$ master and all the individual state trajectories used in its creation, and \emph{ii)} the identified maxima in the master and the corresponding trajectory heights at the associated dates for the individual states. For both cases, we report the averages of these deviations, across all included states, in Section~\ref{sec:3d}.

To characterize improvements afforded by invoking our dynamical clustering, we repeated this process -- creation and compression of a master model, and an evaluation of the errors accrued when na\"ively assuming this archetypal behavior for all the states it summarizes -- separately for each cluster, for both the 446-day and maxima-only cases.

\subsection{Significance of stringency and vaccination trajectory differences among clusters}
\label{sec:2d}

\emph{Averages} of government policy stringency trajectories and vaccination-rate trajectories (introduced in Section~\ref{sec:2a}) were created for each cluster having at least two members by taking the means, across all cluster members, over time. That is, we separated states according to a learned cluster partitioning $\mathcal{C}$ and computed the mean, standard deviation and standard error (of the mean) over the constituent states for each cluster separately, by day.

To develop a framework for discerning whether the $\sim|\mathcal{C}|$ resulting policy (or vaccination-rate) ``average trajectories" exhibited statistically significant differences from one another -- as well as to reduce overfitting and more reasonably approximate sample independence -- we chose not to compare from all daily stringency values, but rather made the approximation that the trajectory values from the \emph{first day of each month} could themselves serve as independent observations on a given cluster-system. We then performed a Krusal-Wallis test, with the number of samples equal to the number $\sim|\mathcal{C}|$ of clusters with at least two state members, drawing observations from the first day of each month between May 2020 and May 2021 (inclusive, so as to associate a total of exactly 13 observations with each group).

As detailed below, vaccination trajectories can be shown to become (visually) distinguishable only after a certain calendar date in the history of the pandemic. We first performed a dynamical clustering above for all vaccination-rate trajectories in the manner described in Sec.~\ref{sec:2b} for infection trajectories; we then created ``average vaccination trajectories" for all the (previously identified) infection-trajectory clusters with at least two states as members, and identified where the vaccination-trajectory slopes were changing by studying their first and second derivatives (built-in ``\emph{diff}" function in Python 3's \emph{NumPy} and \emph{Pandas}). Finally, we performed a Kruskal-Wallis test to determine whether the vaccination-rate values on July 22, 2021 -- a date by which the rates for all states were as well-separated as they would be by the end of our data -- differed significantly, between the aforementioned infection-derived clusters.

We report both test statistics $H$ and $p$-values for $5\%$ significance in Section~\ref{sec:3e}, as well as the results of a post-hoc test (Python implementation of Conover's test as part of the scikit-posthocs package, using ``holm" step-down method with Sidak adjustments to $p$-values) to infer which of the distinct pairs of groups are statistically distinguishable.

\subsection{Relationship between vaccination rates and emerging, new-case rates}
\label{sec:2e}

All the methods discussed to this point have been aimed at establishing that all $n=50$ U.S. daily new infection trajectories fall into a low-dimensional set, $\mathcal{C}$; in Section~\ref{sec:3}, we infer an appropriate value for $|\mathcal{C}|$ (i.e., a partitioning that sufficiently distinguishes coarse differences in ``wave" histories). We also wish to assess the \emph{usefulness} of a cluster set $\mathcal{C}$ for efficiently summarizing U.S. history -- and its potential grounding in reality, beyond its practical use for data compression -- according to how it recapitulates patterns in geography, government stringency, and vaccination rates. 

In Section~\ref{sec:3e}, we see the vaccination-rate trajectories -- percentages of fully-vaccinated individuals represented in each U.S. state's population -- diverge from one another markedly (and then remain divergent) only following the late Spring of 2021. Furthermore, those vaccines against COVID-19 available in the period described by our study -- April 2020 to July 2021 -- require several weeks' time to accomplish even partial seroconversion~\cite{lombardi2021mini,sette2021adaptive}. We therefore studied the degree to which the average vaccination rates within each cluster, \emph{approximately two weeks before the end} of our data, predicted daily-new-case statistics in the corresponding clusters on \emph{the last full day} of data.

We performed tabulations of both the July 22, 2021 vaccination rates and the August 3, 2021 ``daily new infection" records, by cluster, according to the set $\mathcal{C}$ learned and fixed in Section~\ref{sec:3a}. We then investigated the relationship between within-cluster vaccination rates and \emph{i)} the within-cluster \emph{mean} of states' infection trajectory values and \emph{ii)} the within-cluster \emph{standard error of the mean} for these (relative) new-case loads, at an approximate two-week lag, by means of both Spearman rank-order correlations and two linear regressions. For both types of analyses, we considered only states that belonged to clusters encompassing at least two states, omitting ``singleton" clusters that consisted of only a solitary state (see Section~\ref{sec:3a}). We read off the output ``\emph{r}"of the ``\emph{linregress}" function in \emph{Scipy}'s Statistical functions module as the $r^2$ value, or linear (Pearson) correlation coefficient, to further quantify linear trends.

\section{Results}
\label{sec:3}

\subsection{Example data and relevant time frames}

\begin{figure}[H]
    \centering
    \includegraphics[width=\linewidth]{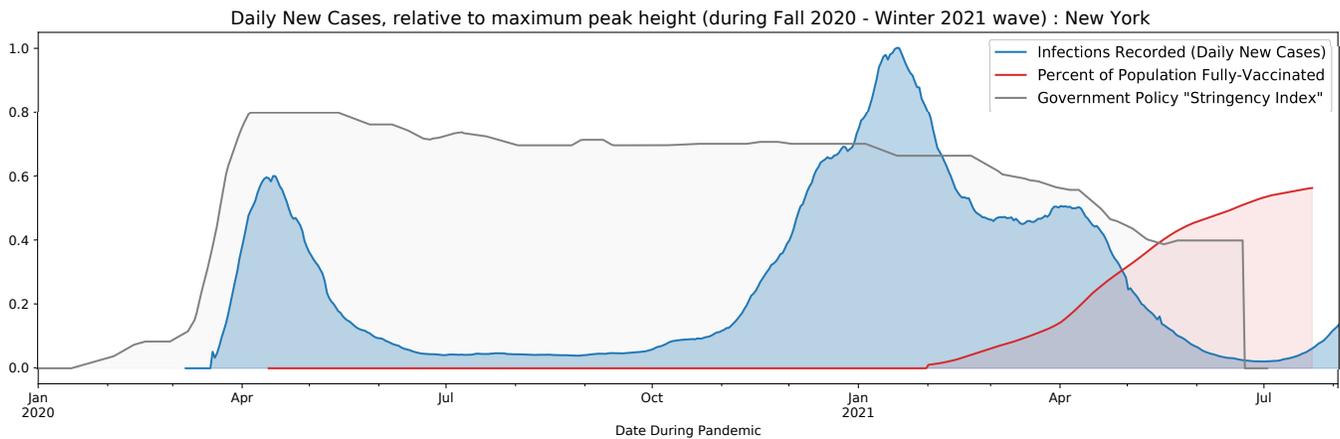}
    \caption{Infection, vaccination, and policy trajectories for the state of New York.}
    \label{fig:trajectorysamples}
\end{figure}

The state of New York exhibited a strong participation in the first, third, and fourth wave opportunities (Fig.~\ref{fig:trajectorysamples}), with essentially no reprieve between the latter two; New York was also among those states exhibiting a relative quiescence during the second wave opportunity. The number of new infections during the fourth wave opportunity (Spring 2021) were comparable to those observed exactly one year prior, during the Spring of 2020. Only the initial weeks of the fifth wave opportunity are depicted, since the infection data source terminates on August 3, 2021.

The vaccination trajectory for New York begins rising after the new-infection peak in the Winter of 2020 - 2021, coincident with the onset of vaccine administration to the public. It changes slope appreciably only during the daily new infections peak that was realized in the context of the fourth wave opportunity (April 2021). Government policy stringency information exists prior to the first wave opportunity, but ends prior to the fifth, in our data sets.

\subsection{Initial clustering and ocular analysis}

\subsubsection{State's unique trajectories fall into 4 major clusters}
\label{sec:3a}

As the cluster threshold is varied from its minimum (0) to its maximum possible value (1), the initial hierarchical clustering of our infection trajectories converges rapidly to resolve just a handful of large clusters. At threshold 0.2, we resolve only 8 multi-state clusters (Fig.~\ref{fig:fig2_clustering}a); approaching threshold 0.43 and beyond, clusters tend to grow in size, while $|\mathcal{C}|$ decreases concomitantly. Once the threshold values reach $0.6$, representing a temporal cross-correlation of at least $1 - 0.6 \rightarrow \mbox{} 40\%$ between any pair of constituent trajectories internal to a cluster, there remain only three big clusters and two outlying states -- California and Hawaii both form their own, standalone, singleton ``clusters."

The 0.43 threshold represented the highest threshold for which non-singleton clusters were still of comparable size. In particular, whereas the three large clusters at the 0.6 threshold are composed of 12, 12, and 24 
states, those at threshold 0.43 have sizes 12, 12, 8, and 16. Although it is possible for the distinct temporal patterns characteristic of the 50 states to fall into differently-sized clusters, we expect that power can be maximized by comparing the statistics of uniformly-sized clusters (the size distribution associated with highest entropy~\cite{cover_book}); by this heuristic, we chose to work predominantly with the clusters resolved at this intermediate threshold value. To visualize salient disparities in the states' temporal dynamics that separate the corresponding, $|\mathcal{C}|=4$ major clusters at this ``locked" threshold, we plot mean infection trajectories for each cluster, color-colored by proximity in correlation-space (Fig.~\ref{fig:fig2_clustering}b); these color-coded data, taken from the training set that ran from April 2020 to July 2021, were superimposed upon gray curves representing the same mean trajectories, but updated Ref.~\cite{jhucrc} to run between March 2020 and August 2021.

\subsubsection{Clusters differ in their wave timings and relative heights}

Commonalities among the mean, cluster-wise trajectories include the aforementioned Winter 2020 - 2021 maxima, as well as the presence of smaller peaks in the Spring and Summer seasons. Although our correlation-based clustering was intended primarily to address the \emph{timing}, or synchronicities, of realized wave opportunities, it also highlights some salient differences in the magnitudes, or \emph{relative heights}, of those opportunities. The distinguishing properties follow.

\begin{figure}[H]
    \centering
    \includegraphics[width=\linewidth]{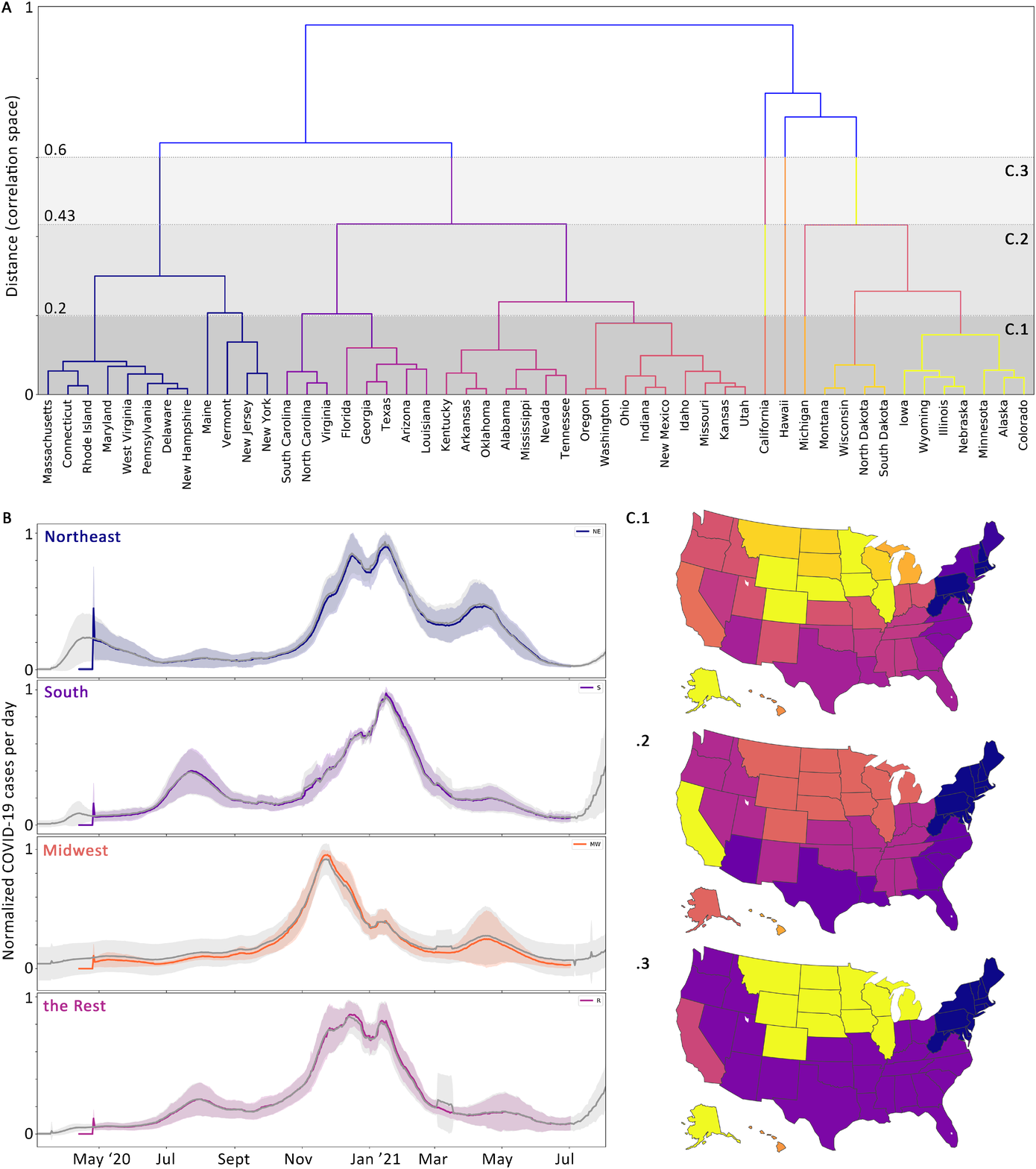}
    \caption{Clustering results for the U.S. state infection trajectories. \textbf{a)} Dendrogram summarizing hierarchical clustering. \textbf{b)} Trajectory mean (standard error) within each cluster; any discrepancies between training data source (color) and testing data (grayscale) are due to the use of independent sources. Outliers (California and Hawaii) are not depicted. \textbf{c)} Geographical distribution of cluster memberships at the different thresholds indicated by gray shading in top panel.}
    \label{fig:fig2_clustering}
\end{figure}

Cluster 1 alone saw a major, initial surge of new cases in Spring of 2020 -- the first wave opportunity -- while the states in Clusters 2 and 3 saw their first collective spike only weeks later, during the second wave opportunity in the Summer of 2020. Cluster 4 is distinguished from Clusters 1-3 by its effective lack of either a Spring or a Summer 2020 surge, with the third, Winter wave opportunity playing host to its first major elevation. At a finer resolution, this principal, Winter maximum is observed to concentrate early (vicinity of December) for Cluster 4, with only a modest resurgence of new infections shortly thereafter; other clusters exhibited a different, more multiply-peaked structure, as well as comparatively \emph{later} peaks, in that season. The closest pair in correlation-space  -- Clusters 2 and 3 -- share most of their coarse features, with aspects of this multiply-peaked Winter setting the latter apart from the former. All U.S. states were observed to have partial resurgences in Spring 2021, 
and every single state experienced an upswing in Summer 2021 (not explicitly depicted), with pronounced variability even \emph{within} clusters by August (Fig.~\ref{fig:fig2_clustering}b).

Collectively, these patterns in the realized heights for each wave opportunity -- that is, the (relatively) medium-sized first wave, essentially absent second wave, and very large third wave expressed by Cluster 1; then, the absent first wave, medium-sized second wave, and very large third wave for Clusters 2 and 3; and then, the absent first and second waves, but large third wave for Cluster 4 -- provide some degree of intuitive explanation for why these particular clusters were resolved. We quantify this intuition by focusing on specific, ``important dates," in Section~\ref{sec:3d}.

\subsection{Correlative and Predictive Cluster Validations}
\subsubsection{Dynamical clusters turn out to be geographically related} 
\label{sec:3b}

Even given the robust set of qualitative dynamical patterns teased out by means of ocular analysis in the previous section, the learned partitioning $\mathcal{C}$ still eludes practical explanation: are there other, non-dynamical patterns associated with the states that were grouped together that recapitulate the learned clustering? In order to visualize the spatial arrangement of those states whose trajectories clustered together at various thresholds, as one such pattern, we colored the states on a U.S. geopolitical map, according to cophenetic similarity (correlation-space distance) as in Fig.~\ref{fig:fig2_clustering}b.

Adjacent, or neighboring, U.S. states were much more likely to be clustered together (Fig.~\ref{fig:fig2_clustering}c). With a few exceptions (including southern states, at the 0.43 threshold, that are separated by land but still connected by water), the majority of the clusters at all three thresholds unite states that lie along relatively uninterrupted, existing geographical paths. As those clusters resolved at small thresholds merge to form larger clusters at larger thresholds, the latter begin to reveal clear correspondences to established U.S. geopolitical regions; these are visible at our ``locked," 0.43 threshold.

In particular, our Cluster 1 predominantly describes states that are traditionally subsumed under ``the Northeast". Cluster 2 describes ``the South," encompassing both the coastal southeastern border and several states along the southern rim of the country. Cluster 4 represents ``the Midwest," only exchanging the states of that region's traditional, southernmost border for the three on its western border. Notably, the two outliers -- California and Hawaii -- are also \emph{geographical} extremes. The remaining Southern, Western, and Central form one, comparably-sized cluster that (with the exception of Colorado) sweeps a clean arc from roughly the latitudinal center of the country to Washington State; this ``Rest" of the states, comprising Cluster 3, is geospatially contiguous (and, at higher thresholds, merges) with its most proximal neighbor in correlation-space, Cluster 2.

It might be reemphasized that all clustering operations were done \emph{dynamically} -- that is, according to correlative similarities between the respective time courses of ``daily new case" loads associated with each U.S. state. No explicit spatial or geographical clustering was considered; rather, any geographical patterns emerged as \emph{consequences} of the dynamical clustering described in Section IIIb.

\subsubsection{Date ranges most explanatory for telling states apart coincide with peak times for whole-U.S. infection ``waves"}
\label{sec:3c}

Analyzing the (cumulative) percentages of variability that can be ``explained" by incorporating exclusively the first $k$ principal components in reconstructions of the daily new infection trajectories (Table I), it is apparent that $k_0=6$ principal components suffice to capture over $90\%$ of the total variance considered by PCA. Reconstructions using the first $k=10$ components produce trajectories almost indistinguishable from the originals (not depicted).

\begin{table}[h!]
\begin{tabular}{cccll}
\hline\hline
\textbf{\begin{tabular}[c]{@{}c@{}}Principal Component\\ Number\\ $k$ \end{tabular}} & \textbf{\begin{tabular}[c]{@{}c@{}}Cumulative\\ \% of Variance\\ Explained\end{tabular}} & \textbf{\begin{tabular}[c]{@{}c@{}}Reconstruction\\ error for first\\ $k$ components\end{tabular}} &  &  \\
\hline\hline
1                                                                                 & .437                                                                                  & .096                                                                         &  &  \\
2                                                                                 & .646                                                                                  & .070                                                                         &  &  \\
3                                                                                & .746                                                                                  & .055                                                                         &  &  \\
4                                                                                & .831                                                                                  & .048                                                                         &  &  \\
5                                                                                & .871                                                                                  & .040                                                                         &  &  \\
6                                                                                 & .908                                                                                  & .035                                                                         &  &  \\
7                                                                                 & .934                                                                                  & .029                                                                         &  &  \\
8                                                                                 & .949                                                                                  & .025                                                                         &  &  \\
9                                                                                 & .962                                                                                  & .022                                                                         &  &  \\
10                                                                                 & .970                                                                                  & .019                                                                         &  & 
\end{tabular}
\caption{Key results of PCA.}
\end{table}

\begin{figure}[H]
    \centering
    \includegraphics[width=\linewidth]{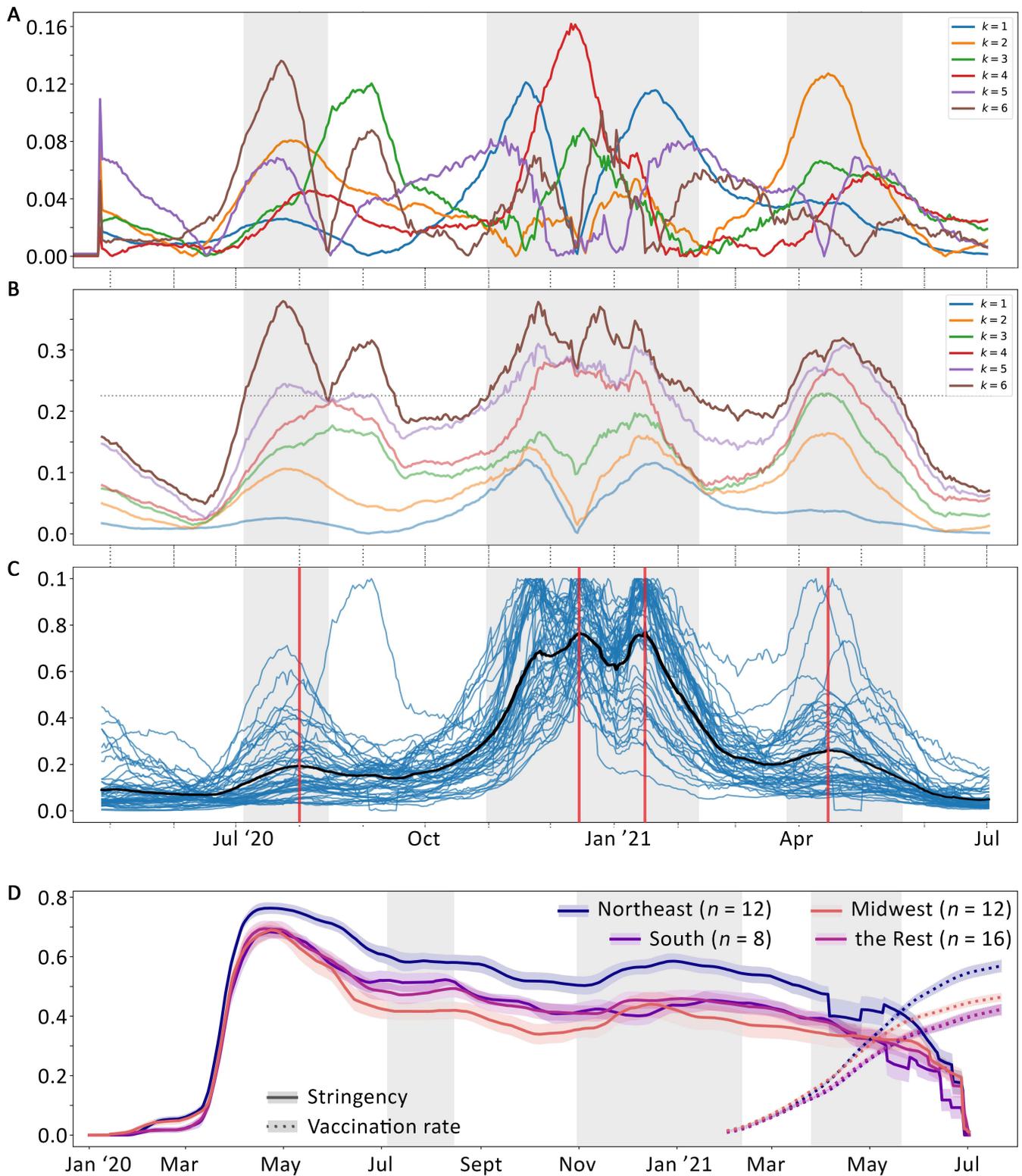}
    \caption{Results of applying PCA to temporal features (daily new infection values) of the $n=50$ U.S. state trajectories. \textbf{a)} Feature importances for individual PCA components. \textbf{b)} Cumulative ($k=1, 
    \dots, 6$), ``overall" feature importances. \textbf{c)} Original $n=50$ U.S. daily new infection trajectories, with master model superimposed in black and ``important" dates for explaining variation as red vertical lines. \textbf{d)} Mean policy and vaccination trajectories, by cluster, for context. Gray boxes indicate regions where daily values of the curve reconstructed from the first $k_0$ eigenvectors exceeded their mean across all encompassed days.}
    \label{fig:fig3_pca}
\end{figure}

Feature importance analysis suggests that the key dates for explaining trajectory variance encompassed long periods from November to December 2020, and January to mid-February 2021, with weaker contributions in the Spring of 2021 and even late Summer 2020 -- that is, all the ``important dates" comprised continuous \emph{ranges} --  for the $k=1$ component (Fig.~\ref{fig:fig3_pca}a). The same is true for the successive components $k\in\lbrace 2, 3, 4, 5 \rbrace$. For instance, the high-prominence ranges for the $k=2$ component coincide with the time-continuous Summer 2020 and Spring 2021 waves. Two lower-prominence peaks in the $k=2$ importance plot describe a more subtle ``fine-tuning" to temporally briefer, Winter fluctuations.

The overall importances imputed for each date, based on the first $k_0=6$ components, also manifest as continuous stretches in time. Above-average values spanned only three, broad date ranges (Fig.~\ref{fig:fig3_pca}b) across the second, third, and fourth wave opportunities. These three ranges all corresponded closely to established date ranges for whole-U.S. peaks in daily new infection records (see discussion of the master curve in Sec.~\ref{mastermodel} below, or Ref.~\cite{policyRef}), implying that the time periods surrounding \emph{global}, national peaks, were also those of the greatest \emph{local}, state-level variation.

\subsubsection{Clustering improves PCA-derived, minimal models of archetypal U.S. infection trajectories}
\label{mastermodel}
\label{sec:3d}

Projecting the $n$ individual state trajectories onto just the first, $k=1$, component resulted in the same, archetypal curve for all states. This curve was identified with the \emph{mean trajectory} over the $n=50$ time series, demonstrating that PCA had ultimately learned to distinguish states by their \emph{deviations} from the collective, average national behavior.

Excluding the Spring of 2020 for the reasons discussed in Sections~\ref{inf_data}~-~\ref{def_waveopps}, all the ``wave opportunities" are represented by maxima in the master curve (red vertical lines in Fig.~\ref{fig:fig3_pca}c). At the prominence level used to detect peaks here (Section~\ref{master_dev}), the multiply-peaked, third wave opportunity splits into two separate maxima for consideration.

Separating the $n$ states by cluster and inferring a master model for each cluster separately recovers the ``average" trajectories for each cluster (see Fig.~\ref{fig:fig2_clustering}b). Reconstructing individual state trajectories using their respective master models can reduce reconstruction error, from at least two perspectives: 1) the compressed, maxima-only representation and the full, 446-day trajectories should be reconstructed more accurately when a master-model construction is done \emph{separately} for each cluster, as opposed to all $n$ states \emph{together} (mean absolute deviations between the states' original and reconstructed trajectories should tend to decrease); 2) the error between an original trajectory and the ``tailored" master made for its containing cluster should diminish, even if no explicit reconstruction is done (the mean absolute deviations between any cluster-averaged trajectory and the original trajectory for given state in the corresponding cluster should be smaller than that between the original trajectory and the full, ``$n=50$" master model). 

Measurements for the latter (Table II), based on either the full, 446-day curves and the 4 key, ``important" dates discerned above confirm that errors are smaller within clusters, but also suggest that encoding the 446-day trajectories as a mere 4-vector of amplitudes at the important dates accomplishes a fair \emph{compression} of states' infection histories: the latter are only slightly higher. Since all error values are of order $\sim 0.1$, and cluster memberships are the only inputs needed to a create master model (beyond the data themselves), specifying the cluster membership -- or, equivalently, its geographical region, according to Fig.~\ref{fig:fig2_clustering}c -- amounts to a prediction of that state's normalized trajectory heights at each wave opportunity with collective, net error of $10\%$ of the height of the corresponding Winter peak, on average.

\begin{table}[h!]
\begin{tabular}{ccccll}
\hline\hline
\textbf{\begin{tabular}[c]{@{}c@{}} Relevant \\ master model \\ for evaluation \end{tabular}} &
\textbf{\begin{tabular}[c]{@{}c@{}} Error (MAD), \\ based on full \\ master model \end{tabular}} &
\textbf{\begin{tabular}[c]{@{}c@{}} Error (MAD),\\ based on peaks \\ of master model \end{tabular}} & 
\textbf{\begin{tabular}[c]{@{}c@{}} Difference in \\ (MAD) error, \\ full $\rightarrow$ peaks \end{tabular}} &  & 
\\
\hline\hline
                                        
All states ($n=50$)             & .0962     & .1514     & .0552     &  &  \\
Cluster 1 ($n=12$)              & .0624     & .0933     & .0309     &  &  \\
Cluster 2  \mbox{} ($n=8$)      & .0514     & .0694     & .0180     &  &  \\
Cluster 3 ($n=16$)              & .0611     & .0882     & .0271     &  &  \\
Cluster 4 ($n=12$)              & .0562     & .0906     & .0344     &  & 
\end{tabular}
\caption{Errors associated with average, state-wise deviation from master models}
\end{table}

\subsubsection{Government stringencies \& vaccination rates for different clusters have different average behaviors over time}
\label{sec:3e}

The geographical considerations above (Fig.~\ref{fig:fig2_clustering}c) support the notion that there might be ``real-world" correspondences to the four, abstracted clusters identified in Sec.~\ref{sec:3a}. A real-world relevance for these clusters is further supported by their ability to find discriminatory patterns in two other metrics, neither of which was included in their definition: government policy stringency scores and vaccination rates (shown for temporal context in Fig.~\ref{fig:fig3_pca}d; see Section~\ref{ancillary}).

For all Clusters 1 - 4, the Spring of 2020 begins with a rapid increase in stringency that reaches an all-time high around April, despite the fact that not all states realized an equally strong ``first wave" opportunity (see Fig.~\ref{fig:fig2_clustering}b). It is clear that \emph{i)} average stringency scores are highest for the Northeast, and lowest within the Midwest, through time; \emph{ii)} the remaining clusters are less distinct from each other than they are from Cluster 1; \emph{iii)} Clusters 2 and 3, the most overlapping, sustained stringencies \emph{intermediate} to those of the Northeast and Midwest for the majority of time.

Repeating this analysis for the vaccination-rate trajectories again highlights the Northeastern states as the highest-scored cluster, and reemphasizes that high degree of similarity between the Southeastern rim and the ``Rest" of the Southern, Central and Northwestern regions that we had initially established via infection-trajectory clustering. In contrast to the overall pattern observed for stringencies, these latter groups are the lowest-ranked for vaccination.

Statistical hypothesis testing (four-sample Kruskal-Wallis, with 13 observations in each group) suggests that the monthly values for the mean stringency scores ($H=17.6, ~ p=5\cdot10^{-4}$) and vaccination rates ($H=31.9, ~ p=5\cdot10^{-7}$) are \emph{not} identical across the four clusters. Adjusted $p$-values for the various, possible pairings suggest that Cluster 1 was distinct from both Clusters 2 and 3 ($p=2\cdot10^{-2}$) as well as from Cluster 4 ($p=5\cdot10^{-5}$), but that Clusters 2 and 3 were statistically indistinguishable from each other ($p=9\cdot10^{-1}$), and possibly even from Cluster 4 ($p=1\cdot10^{-1}$).

The vaccination trajectories can be observed to diverge appreciably only around the start of April, and even then a hierarchical, correlation-based clustering (Section~\ref{sec:2b}) unifies all $n=50$ trajectories at a ``distance" of $\sim 0.025$, a full order of magnitude smaller than that of the most tightly-correlated merge (Cluster 2 in Fig.~\ref{fig:fig2_clustering}b) among the infection trajectories. Also, numerical differentiation (not depicted) reveals that the only major, dramatic slope changes occur around April 1 and May 1, 2021, so that later divergences are largely explained via higher vaccine ``adoption rates," on average, in the early stages of availability among states belonging to certain clusters. All this evidence for a ``low distinguishability" through the Spring of 2021 and an apparent separability only later, in Summer 2021, was consistent with the Kruskal-Wallis test for cluster-wise differences among July 22, 2021 vaccination rates: $H=20.6, ~ p=1\cdot10^{-4}$.

Among the four clusters containing at least two states each, Conover's post-hoc tests suggests that Cluster 1 alone was unambiguously distinguishable from all the others (Cluster 2, $p=2\cdot10^{-4}$; Cluster 3, $p=1\cdot10^{-5}$; Cluster 4, $p=9\cdot10^{-3}$), and that Clusters 2 and 3 were completely indistinguishable from one another ($p=0.968$). Post-hoc testing did not indicate strong degrees of separation between Cluster 4 and Cluster 2 ($p=0.234$) or Cluster 3 ($0.170$).

The fact that vaccination-rate trajectories do not differ significantly until the concluding region of our time series  indicates that the only vaccination values potentially useful in discriminative \emph{predictions} are those from Summer 2021.

\subsection{Seatbelt Hypothesis: Within-cluster vaccination rate predicts uncertainty in future new-case upticks}
\label{vacc_std}
\label{sec:3f}

\begin{figure}[H]
    \centering
    \includegraphics[width=\linewidth]{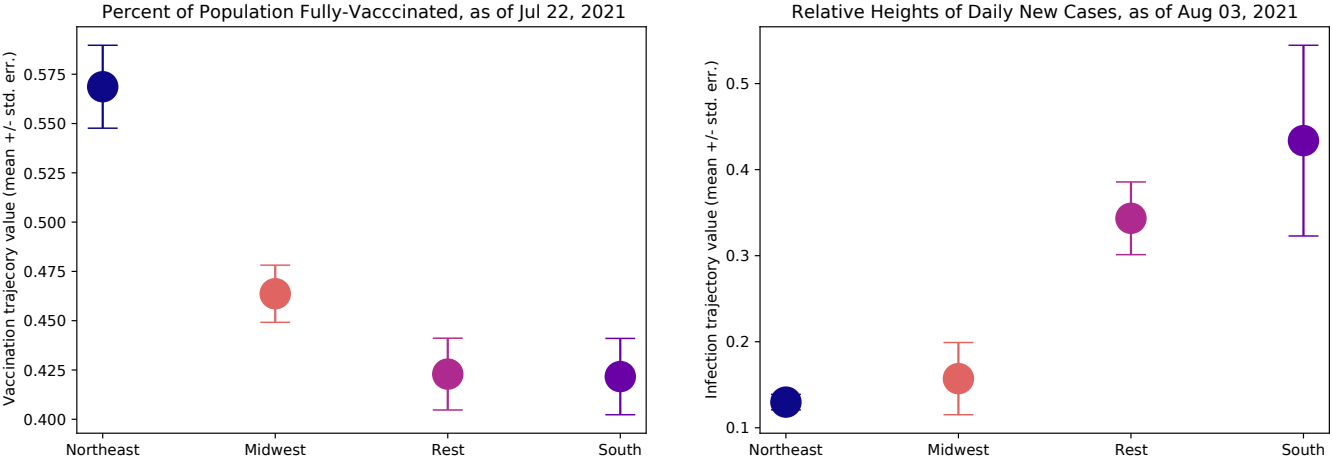}
    \caption{``Seat belt" hypothesis: organizing the clustered regions in order of decreasing vaccination rate reveals negative relationships with the mean and standard error of new, emerging cases rates, at a two-week, lag, during Summer 2021.}
    \label{fig:seatbelt_dnc_plus_vacc}
\end{figure}

Correlation and regression analyses suggest monotonic relationships between our cluster mean vaccination rates and several aspects of the ``average infection trajectories" associated with the corresponding clusters, at a two-week lag. First, the mean values of the daily new infections in any given cluster at the beginning of August, relative to the corresponding Winter 2021 peak values, are inversely related to the mean vaccination rate in that cluster (Spearman rank correlation coefficient $-1$; Fig.~\ref{fig:seatbelt_dnc_plus_vacc}). Emphatically, this relationship is not strictly linear. For instance, several Midwestern states -- the cluster with the \emph{second-lowest} mean in daily new cases in early August -- recorded fewer cases than all the Northeastern states, the cluster with the \emph{lowest} mean in daily new cases in August. Still, the Pearson correlation coefficient associated with the linear-order, pairwise correlation between mean vaccination rates and mean case loads at a two-week lag is measured (via regression or direct computation) as $-0.8$, suggesting a directional trend.

Beyond the means, the average, within-cluster July 22 vaccination rate also closely tracks another quantity -- namely, the \emph{standard error} of the mean in the daily new infection trajectory values at the beginning of August. Specifically, the fraction of fully-vaccinated individuals is proportional to the ``spread" of -- or, ``uncertainty" in -- trajectory values realized across the clusters' constituent states. Concretely, for the most \emph{highly-vaccinated} group of states (Cluster 1, the Northeast) the range of observed infection-trajectory upticks is the \emph{smallest}, and this range tends to increase with decreasing vaccination coverage (Spearman rank correlation coefficient of $-1$). Fitting a linear model to measure the Pearson correlation coefficient (value $-0.74$) reveals a strongly negative trend: vaccination rates recorded in late July are \emph{positively} correlated with a measure of case-load predictability in early August (not depicted in Fig.~\ref{fig:seatbelt_dnc_plus_vacc}).

Consider, as an analogy, the adoption of seat belts as a measure of protection against severe injury. Seat belts do not insure against the acquisition of all wounds, but are designed to be good at decreasing the probability of fatalities. One cannot guarantee survival in the worst of accidents by using a seat belt, nor is one ever bound to acquire severe injuries when failing to activate the device -- yet the \emph{range} of the appreciably likely injuries that one might receive, in the context of common accident configurations, can be constrained via wearing a seat belt. Similar is the pattern suggested by Fig.~\ref{fig:seatbelt_dnc_plus_vacc}: among those states belonging to clusters for which vaccination was more widespread in late July, the \emph{range} of relative trajectory heights two weeks later, at the beginning of August, is observed to be smaller. In that short time period during which vaccination trajectory values differed significantly among state clusters, these values were also predictive of yet unseen, ``future" infection rates.

\section{Discussion}

Here, we demonstrate that dynamical clustering allows separation of all 50 states into a small number of clusters (we settled on four, with two outliers). Meanwhile, Principal Components Analysis corroborated the existence of several distinct ``waves" of COVID-19 infections nationally. These encompassed four key dates, which served in to build accurate compressions of cluster ``average trajectories" with little other information, showing that a low-dimensional model can capture much of the variation in relative COVID-19 case loads over time.

The fact that our set of clusters $|\mathcal{C}|$ happened to fall into established geographical regions, and showed correlations to state stringencies and vaccination rates further supports the potential for the clusters we discovered to be accurately reflecting reality. Models trained using states' data through June 2021 were useful in predicting standard deviations of cluster case loads in the subsequent, ``fifth wave" from July through September 2021. One parsimonious explanation is that the vaccination rates do correlate with case loads, but that greater usefulness lies in their (positive) correlation with the \emph{predictability} of case loads in our period of interest. Based on all these results, we developed the ``seat belt" hypothesis, founded on the comparison that, while seat belts don't prevent accidents, they do restrict the likely range of adverse outcomes. In this vein, high vaccination rates may not prevent COVID-19 from affecting a state, but states with lower vaccination rates might expect a higher potential \emph{ceiling} for their daily new case loads.

Our results are consistent with previous analyses that identified four, historical waves~\cite{policyRef}, and a role for geography in identifying states with analogous pasts~\cite{NYtimesGeog}, although our findings recapitulated both without explicit geographic input to our models. Nevertheless, there are several important considerations for interpreting our own results in comparison with other analyses. First, we have normalized all case rates to one peak at the end of 2020, so that our results relate to \emph{relative change} and not \emph{absolute case loads}. That is, we clustered states not based on how many cases they have recorded, but the synchronous changes-of-shape in their infection patterns over time. Additionally, we developed models only of reported \emph{cases}, and different results might have arisen had we focused on other metrics, as in deaths or hospitalizations. Similarly, many more data sources could conceivably be added to this model and might change the specific predictions. We chose not to pursue a more exhaustive data-scraping approach, instead choosing to focus on the capacity to make useful predictions with a simple model without either wide-spread exploration or explicit notions of causality, both of which could soak up arbitrary amounts of future time and energy.

As yet more new data becomes available, it will be possible to ``test" the hypothesis that higher vaccination rates will translate to a restricted ``likely range" of new case loads generally. Here, too, future researchers must exercise caution. For example, whereas our vaccination-rate trajectories remain similar over long stretches of time, a more exhaustive analysis would have to rule out that infection dynamics do not bifurcate into qualitatively different behaviors if local vaccination rates reach a certain (even sub-herd immunity) value. On this note, we emphasize that our state- and state cluster-level results might be superseded by different insights at, for instance, the level of U.S. counties. In addition to these considerations, as well as the overwhelming likelihood that vaccination prevalence alone does not control new-infection dynamics, there remains the possibility that other complex influences -- such as geographic spread~\cite{fauci_talk_Sep2020} or climate patterns~\cite{bukhari2020effects} could conceivably translate to different clusters ``taking the lead" in emerging, new-case rates at different times. We stress that our ``seat belt" measurements were evaluated while U.S. states were experiencing a clear upswing, ostensibly due to the spread of a novel SARS-CoV-2 variant.

We emphasize again that, while our models do seem to provide useful predictions and to correspond to other real-world features, they are not causal models. The fact that state' dynamics can be identified and tied into clusters with few inputs is efficient, but does not resolve the issue of \emph{why} compact patterns exist. Explorations along these lines do remain important, and we hope our model might be useful enabling others to add more variables in search of causality. Effectively infinite other data sources could be explored in this endeavor, but given the behavior of the clustering we observed, some possibilities suggest themselves, including mask-use policies and climate. While we do not assess political data~\cite{githubCsEsetc.} explicitly in our model, the regional clustering and relative differences in stringency suggest that they maybe worth exploring in these contexts as well.

While our pipeline is not the only existing approach to compressing infection-rate information in such a way that is useful for prediction, it offers distinct advantages. Several analyses elaborated previously, as in Ref~\cite{policyRef}, focused predominantly upon tracking and categorizing variations in state-level policies, and provided the basic terminology for describing the three completed ``waves" of COVID-19 cases in the United States (as well as ``concerns of a fourth wave" at the time of its publication in Spring 2021)~\cite{policyRef}. This four-wave structure was discerned from observations on the overall patterns of COVID-19 cases across the U.S (the sum of contributions from all 50 states); instead, we \emph{impute} a data-driven number of waves, and find that the encompassed dates agree well with those of Ref.~\cite{policyRef}.

This latter sentiment is also echoed within two earlier studies that had explored multiple clustering methods~\cite{james2020cluster} and attempted to classify U.S. states as being within their first or second ``surge"~\cite{james2020covid}. Unlike our work here, the ``optimality" of a given partitioning $\mathcal{C}$ in Ref.~\cite{james2020cluster} is not evaluated on whether clusters predict external, ancillary variables, but by analytic means. The analyses in Ref.~\cite{james2020covid} also focused on ``daily new cases," which we chose over other metrics, such as cumulative cases or daily death rates, as a snapshot of the evolving transmission situation within the U.S. The authors of the present work were unaware of both works while conducting the research described here, but our contribution can nonetheless be considered an extension or generalization of the former: aside from the benefit of a somewhat more retrospective look at the progression of the pandemic, our emphasis on external validation provides for more direct application of clustering to future decisions and research.

In conclusion, our model supports the idea that gross fluctuations the U.S. states' COVID-19 case rates are more regionally coordinated than is revealed by a focus on raw case numbers alone. As a result, while others have reported occasions on which California and Florida may have seen similar statistics~\cite{healthline,washpost}, our findings suggest that a neighbor like Georgia would be more likely to provide actionable information for what is likely to happen to Florida (especially in the context of a new wave) than would California or New York. It remains to be shown, in future work, whether these same states would unite and cluster together on other grounds (e.g., demographic, political, or climatological), and if the specific groupings discovered here serve useful in future infectious disease containment or mitigation strategies.

We have demonstrated that the fifty U.S. ``wave" histories fall into just four distinct infection patterns, and that each pattern can be associated with a specific geographical region. Grouping states by infection pattern revealed that within-group vaccination rates predict key new-case statistics during the early stages of the Summer 2021 surge.

\begin{acknowledgments}

The authors would like to extend a special thanks to the CaliBaja Center for Resilient Materials and Systems, and to those at the San Diego Supercomputer Center, for hosting the ENLACE 2021 program.

\end{acknowledgments}

\bibliographystyle{unsrtnat}
\bibliography{bibliography.bib}

\end{document}